\documentclass{aa} 
\usepackage{txfonts}
\usepackage[utf8]{inputenc}

\usepackage{graphicx}
\usepackage{newtxtext}

\usepackage{ae,aecompl}
\usepackage[colorlinks,citecolor=blue, menucolor=blue, urlcolor=blue, linkcolor=blue]{hyperref}
\usepackage{url}

\usepackage{amsmath}	
\usepackage{amssymb}	
\usepackage{gensymb}

\newcommand{\Msun}{ h^{-1}{\rm M_{ \odot}}}
\newcommand{\hMpc}{ h^{-1}{\rm Mpc}}

\newcommand{\ihMpc}{ h\,{\rm Mpc}^{-1}}
\newcommand{\mycomment}[1]{}

\newcommand{\code}{\tt{}}

\newcommand{\orcid}[1]{\href{https://orcid.org/#1}{\includegraphics[width=10pt]{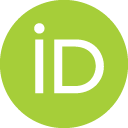}}}

\begin{document}

\title{Baryonification extended to thermal Sunyaev Zel'dovich}

\author{Giovanni Aric\`o$^{1,2}$  \orcid{0000-0002-2802-2928} 
\thanks{E-mail: \href{giovanni.arico@uzh.ch}{giovanni.arico@uzh.ch}}}
\author{Raul E. Angulo$^{2,3}$}

\author{Giovanni Aric\`o
          \inst{1,2}\fnmsep\thanks{E-mail:giovanni.arico@uzh.ch (GA)} \&
         Raul E. Angulo\inst{2,3}}
 
\institute{
Institut für Astrophysik (DAP), Universität Zürich, Winterthurerstrasse 190, 8057 Zurich, Switzerland \and
Donostia International Physics Center (DIPC), Paseo Manuel de Lardizabal, 4, 20018, Donostia-San Sebasti\'an, Guipuzkoa, Spain\and
IKERBASQUE, Basque Foundation for Science, 48013, Bilbao, Spain}

\date{Received XXX; accepted YYY}

\abstract{ Baryonification algorithms model the impact of galaxy formation and feedback on the matter field in gravity-only simulations by adopting physically motivated parametric prescriptions. In this paper, we extend these models to describe gas temperature and pressure, allowing for a self-consistent modelling of the thermal Sunyaev-Zel'dovich effect, weak gravitational lensing, and their cross-correlation, down to small scales. We validate our approach by showing that it can simultaneously reproduce the electron pressure, gas, stellar, and dark matter power spectra as measured in all BAHAMAS hydrodynamical simulations. Specifically, with only two additional free parameters, we can fit the electron pressure auto- and cross-power spectra at 10\% while reproducing the suppression in the matter power spectrum induced by baryons at the per cent level, for different AGN feedback strengths in BAHAMAS. Furthermore, we reproduce BAHAMAS convergence and thermal Sunyaev Zel'dovich angular power spectra within 1\% and 10\% accuracy, respectively, down to $\ell=5000$. When used jointly with cosmological rescaling algorithms, the baryonification presented here will allow a fast and accurate exploration of cosmological and astrophysical scenarios. Therefore, it can be employed to create mock catalogues, lightcones, and large training sets for emulators aimed at interpreting forthcoming multi-wavelength observations of the large-scale structure of the Universe.}


\titlerunning{Baryonification extended to tSZ}
\authorrunning{G.Aricò \& R.E.Angulo} 

\maketitle

\begin{figure*}
\centering
\includegraphics[width=.9\textwidth]{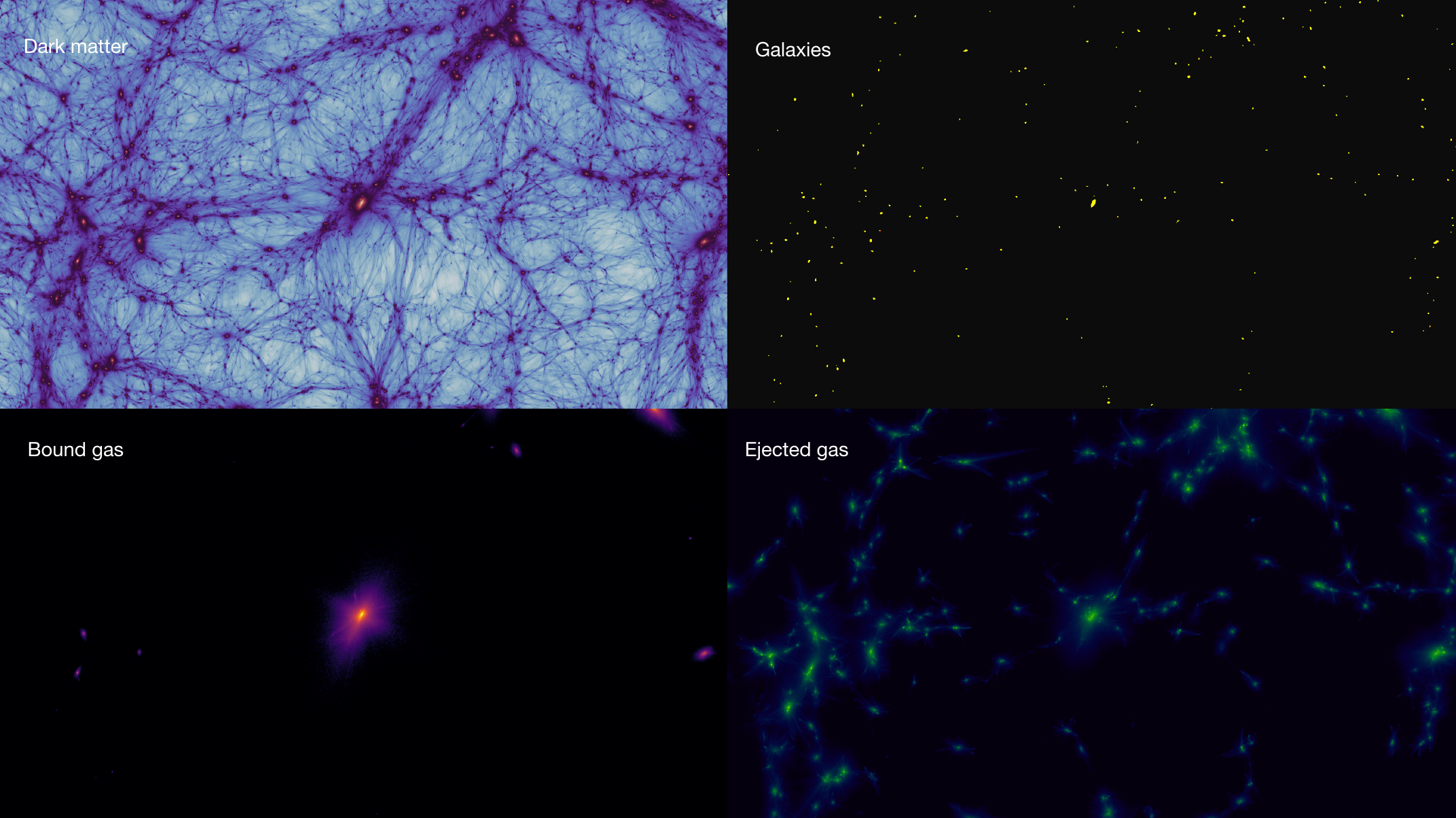}
\caption{Cosmic density fields obtained by applying our baryonification to an $N$-body simulation. Specifically, the total matter density field is decomposed into dark matter (top left), galaxies (top right), bound gas in hydrostatic equilibrium (bottom left), and gas ejected by feedback processes (bottom right). The BCM parameters have been set to values that generate strong AGN feedback.}
\label{fig:cosmic_fields}
\end{figure*}

\section{Introduction}
\label{sec:intro}

The increasing precision of modern Large-Scale Structure (LSS) surveys has pushed the statistical errors of the measurements below the level of uncertainties of our theoretical models. To fully exploit the potential of current and future surveys, an overall improvement in the accuracy of our models as well as an understanding of systematic errors is required. 

For instance, in the analysis of weak lensing (WL) data, a large fraction of data cannot be used because of a dominant systematic theoretical error; the unknown effects of baryonic processes at small scales \citep[e.g.][]{Krause2021,Secco2022,Amon2022}. 
Recently, baryonification algorithms \citep{Schneider&Teyssier2015,Schneider&Teyssier2019,Arico2020, Arico2020c} have been used to flexibly model the relevant baryonic processes and to robustly infer the cosmological parameters \citep{Schneider2021,Arico2023,Garcia2024}. However, the marginalisation over several unconstrained free baryonic parameters degrades the cosmological information that can be extracted, and might introduce prior volume effects in the analyses. In this regard, the simultaneous analysis of multiple observables and their cross-correlation is particularly interesting because it allows the removal of independent systematics and the breaking of parameter degeneracies. 

By informing the baryonification with external data sets, we can better constrain the baryonic parameter space and break degeneracies with cosmology. This approach was pursued, e.g. by \cite{Schneider2021} and \cite{Bigwood2024}, where the authors combined cosmic shear and kinematic Sunyaev-Zel'dovich (kSZ) data, plus the gas fraction from X-ray data in the case of \cite{Schneider2021}. \cite{Grandis2023}, instead, used a combination of X-ray gas fractions and X-ray gas profiles to predict the suppression in the matter power spectrum caused by baryons. 

Another possible avenue is the exploitation of weak-lensing cross-correlations with thermal Sunyaev-Zel'dovich (tSZ). These two probes are highly complementary; the first one provides the integral along the line of sight of matter, and the second one provides the electron pressure. Baryonic physics impacts the two observables with different angular and redshift dependencies. Various analyses of the cross-correlation between WL and tSZ were carried out in the last decade \citep{VanWaerbeke2014,Hojjati2017,Osato2020c,Gatti2021,Pandey2021,Troster2021}. Despite the improvements in the theoretical modelling, the limiting factors are the lack of tight priors on the nuisance parameters and the difficulties of modelling the non-linearities and baryonic physics.

Currently, the most self-consistent predictions of the (non-linear) WL x tSZ cross-correlations come from cosmological hydrodynamical simulations. These simulations simultaneously follow the gravitational collapse of structure and the evolution of gas thermodynamics, using numerical prescription to implement galaxy formation and astrophysical feedback. However, given their computational costs, reaching the required volumes and resolutions is difficult. Another complication is that we need predictions covering a large parameter space in cosmology and astrophysics. Modern supercomputers and codes are filling up this gap, increasingly larger simulations can be run, and the calibration of the so-called subgrid parameters is improving as well \citep{Schaye2010, LeBrun2014, McCarthy2017, vandaalen2020, Antilles2023, Hernandez2023, Ferlito2023, flamingo1, flamingo2}, but it is still not sufficient for full Bayesian analyses. Moreover, it is not guaranteed that the physical models implemented in these simulations are correct.

Alternatively to hydrodynamical simulations, one can extend the halo model to model baryonic components empirically \citep[e.g.][]{Mead2020b, Pandey2024}; or paste baryons into $N$-body simulations, either with machine learning \citep[e.g.][]{Troester2019}, or with analytical profiles  \citep[e.g.][]{Osato&Nagai2022}.

In this work, we take a different path, building on the Baryon Correction Model \citep[BCM, a.k.a. baryonification,][]{Schneider&Teyssier2015,Schneider&Teyssier2019,Arico2020,Arico2020b}. This algorithm perturbs the mass field of $N$-body simulations taking into account the effect of several baryonic components. Taking advantage of the baryonification framework, which provides the gas density profiles of haloes in the $N$-body simulation, we compute analytically the expected temperature and pressure profiles, under the assumptions of a polytropic equation of state for the gas. In this way, we can provide accurate modelling of both dark matter and baryonic fields up to small scales, and allow for joint predictions of WL and tSZ. We validate our approach by comparing it against a state-of-the-art hydrodynamical simulation, BAHAMAS \citep{McCarthy2017,McCarthy2018}. 

Our baryonification approach inherits the benefits of the models presented in \cite{Mead2020b,Pandey2024,Osato&Nagai2022}, and addresses some of their shortcomings. For example, it does not lack accuracy in the transition between 1-halo and 2-halo terms as the halo model, and does not rely on the assumption of linear bias. It also ensures a better consistency between mass and pressure profiles than the baryon pasting algorithm. Additionally, our method does not require hydrodynamical simulations for calibration and is self-consistent at the field level, thus it can be used for predicting higher-order statistics.

This paper is structured as follows:
in \S\ref{sec:model} we summarise the baryon correction model and show how to extend it to thermodynamical properties; in \S\ref{sec:tsz} we explain how to predict the tSZ from a baryonified simulation; in \S\ref{sec:comparison} we compare against the BAHAMAS hydrodynamical simulation; in \S\ref{sec:conclusions} we draw our conclusions. 

\section{Modelling}
\label{sec:model}
In this section, we briefly recap the baryonification, and explain the steps needed to extend the algorithm to predict thermodynamical properties. 

\subsection{Baryonification}
Baryonification algorithms displace the particles in $N$-body simulations to account for the effects of stars and gas on the dark matter \citep{Schneider&Teyssier2015}. For each halo of an $N$-body simulation snapshot, the density profiles of several baryonic and dark matter components are analytically calculated, and then summed to obtain a total "baryonified" density profile. From the difference between this and the original "gravity-only" halo profile, one can work out the displacement field that, when applied to the halo particles, corrects for baryonic effects.

With slightly different flavours, the baryonification has been proven to be accurate in modelling the 2-point \citep{Schneider&Teyssier2015,Schneider&Teyssier2019,Schneider2020,Arico2020, Arico2020c,Giri2021} and 3-point \citep{Arico2020c} statistics of the cosmic matter field, WL peaks \citep{Weiss2019,Lu&Haiman2021,Lee2022}, galaxy-galaxy lensing \citep{Contreras2023,Chaves2023}, X-ray gas fraction and density profiles \citep{Grandis2023}, and it was applied to cosmic shear data \citep{Schneider2021, Chen2023, Arico2023, Garcia2024,Bigwood2024}, lensing maps \citep{Lu2022,Fluri2022,Kacprzak2023,Lu2023,Gatti2024}, and cross-correlations of dark matter haloes with cosmic dispersion measures \citep{Masato2022}. 

\subsection{Multiple cosmic fields}

A standard baryonification only predicts the overall cosmic density field, without specifying the individual distributions of gas and stars.

In App. B of \cite{Arico2020} it was proposed to subsample the matter field, by tagging each particle as part of a given baryonic component and reassigning its mass. In this way, it is possible to end up with separate cosmic density fields for dark matter, gas, and stars. 

Here we take a similar approach with minor modifications. The methodology proposed in \cite{Arico2020} has a relatively high level of discreteness noise, caused by the subsampling of the initial number of particles. In particular, the more baryonic components used, the fewer particles available for the sampling. Here, we avoid this problem by splitting each particle in the baryonified simulation into $N_{\rm c}$ particles, where $N_{\rm c}$ is the number of BCM components considered. These $N_{\rm c}$ particles have the same positions as the parent particle, and their masses sum up to the parent particle's original mass. 

To do this, we take the differential cumulative halo mass profiles of the BCM components $\Delta M_i^j=M_i^{j}-M_i^{j-1}$, and the cumulative final mass profile $\Delta M_{\rm BCM}^j = \sum_{i=0}^{N_{\rm c}} \Delta M_i^j$, where the index $i$ runs over the baryonic components and the index $j$ over the halo radial bins. The mass of each particle inside $r_{200}$ (prior to the BCM displacement), for a given component and radial bin, will be:

\begin{equation}
m_i^j = \Delta M_i^j / \Delta M_{\rm BCM}^j \times m_{\rm GrO}
\end{equation}
where $m_{\rm GrO}$ is the original mass of particles in a gravity-only simulation. One can decide to create a different cosmic field for each individual BCM, or to group some components together (e.g., a component summing bound and ejected gas, or one summing central and satellite galaxies). For the particles outside $r_{200}$ (prior to the BCM displacement), we simply fix the mass of dark matter particles as $m_{\rm DM} = (1-\Omega_{\rm b}/\Omega_{\rm m}) \times m_{\rm GrO}$, and the one of the gas as $m_{\rm gas} = \Omega_{\rm b}/\Omega_{\rm m} \times m_{\rm GrO}$. 

In Fig. \ref{fig:cosmic_fields} we show a visualisation of the cosmic density field for dark matter, galaxies, bound and ejected gas, as generated by baryonifying one of the simulations described in \cite{Arico2020} with the algorithm presented here. Since we have opted in this case for strong AGN feedback, only the most massive haloes in the simulation have retained a gas reservoir, while small haloes have their gas ejected into the intracluster medium, tracing fairly well the large-scale structure. Galaxies are placed at the centres of their host haloes, and their size and mass depend on the masses of the respective host haloes, according to the relations detailed in \cite{Arico2020}.  

\begin{figure}
\centering
\includegraphics[width=.48\textwidth]{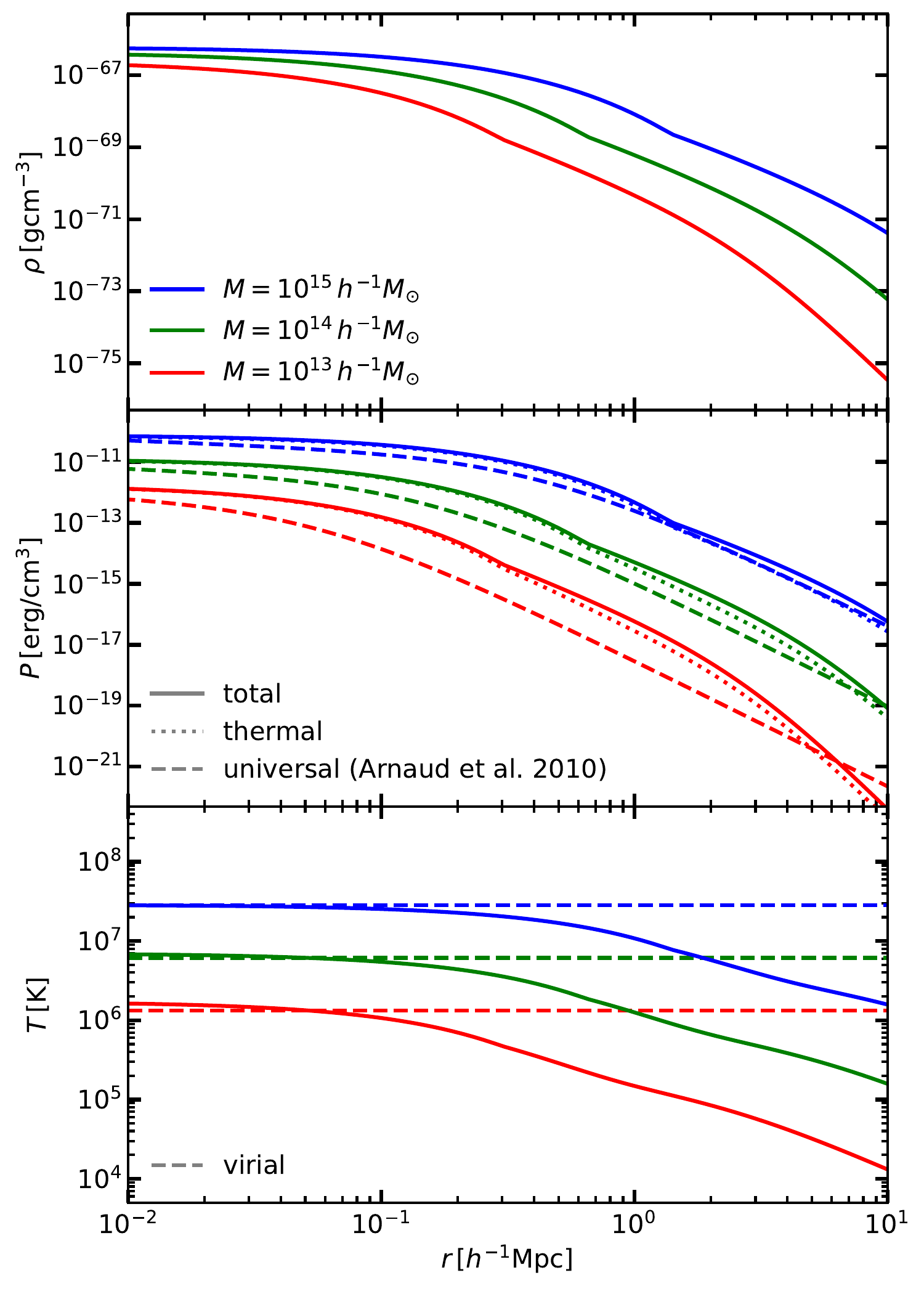}
\caption{Density (upper panel), pressure (central panel), and temperature (lower panel) profiles of the bound gas for haloes in the mass range $10^{15}$-$10^{15}$ ${\it h}^{-1} M_{\odot}$, according to the legend. We show both total and thermal pressure, with solid and dotted lines, respectively. For visual comparison, we show with dashed lines the ``universal'' pressure profiles from \protect\cite{Arnaud2010} and the virial temperature. We stress that the BCM parameters allow for changes in the shape and normalisation of the profiles, and in this case, we did not tune them to reproduce any specific profile.}
\label{fig:bound_profiles}
\end{figure}

\subsection{Thermodynamical profiles}

\begin{figure*}
\centering
\includegraphics[width=.49\textwidth]{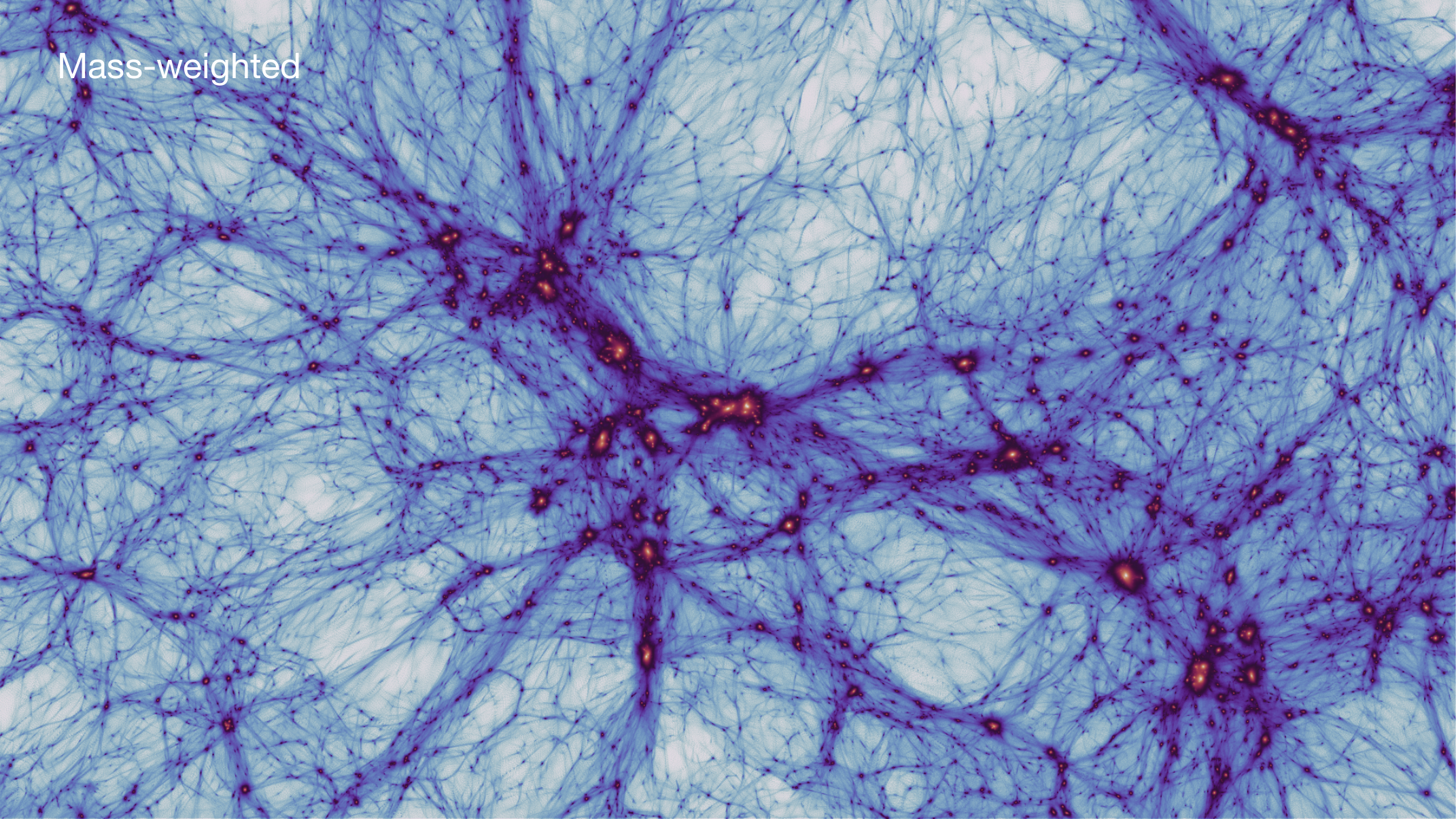}
\includegraphics[width=.49\textwidth]{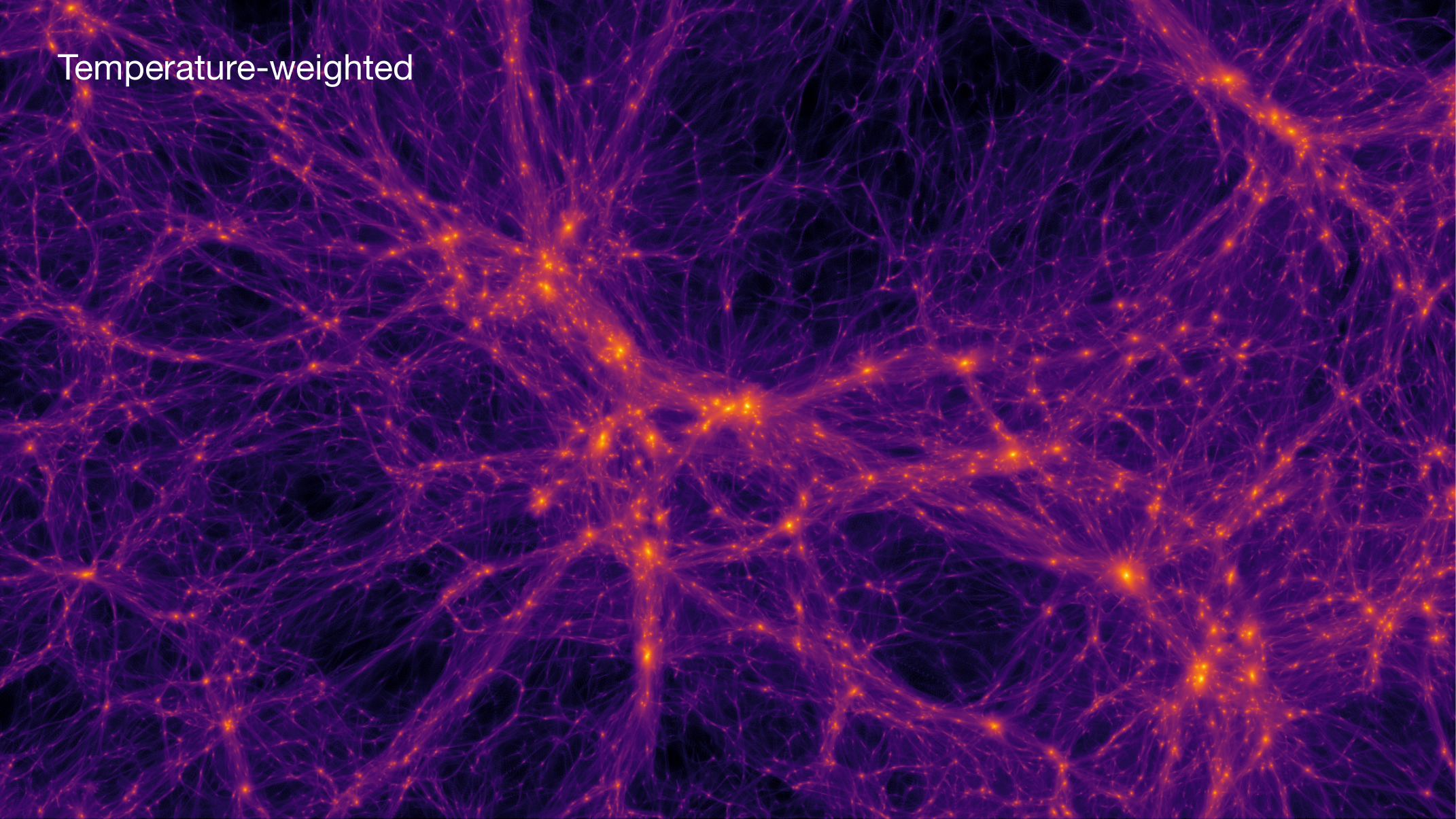}
\caption{Density field in a baryonified simulation, weighted by gas mass (left panel) and temperature (right panel). The BCM parameters have been tuned to generate relatively strong AGN feedback.}
\label{fig:bar_fields}
\end{figure*}

To model the thermodynamical state of the gas, we first parameterise pressure and temperature profiles in haloes, and then assign to each particle a given pressure and temperature, according to its radial distance from the halo centres. 

We start from the analytical gas density profile employed in \cite{Arico2020b}, 

\begin{equation}
\rho_{\rm BG}(r\le r_{\rm out}) = \rho_0 \frac{1}{ (1+r/r_{\rm inn})^{\beta_{i}} } \frac{1}{(1+(r/r_{\rm out})^2)^2},
\label{eq:rho_bg}
\end{equation}

\noindent where $\rho_0$ is a normalisation factor,  $r_{\rm inn}$ and $r_{\rm out}$ define where the slope changes at small and large radii, respectively. We define the inner radius $r_{\rm inn}=\theta_{\rm inn} \times r_{200}$ and outer radius $r_{\rm out}=\theta_{\rm out} \times r_{200}$, beyond which the gas profile perfectly traces the dark matter, i.e. it is described by a Navarro-Frenk-White (NFW) profile \citep{NFW1997}. The gas inner slope depends on halo mass as
\begin{equation}
\beta_{i}=3-(M_{\rm inn}/M_{200})^{\mu_{i}}. 
\end{equation}
We set $\theta_{\rm inn}$, $\theta_{\rm out}$, $M_{\rm inn}$, and $\muß_{i}$ as free parameters. 

Then, we assume a polytropic equation of state of the gas, i.e. that the total gas pressure follows $P_{\rm BG} \propto \rho_{\rm BG}^{\Gamma}$, where $\Gamma$ is the polytropic index. 
Specifically, we set 
\begin{equation}
P_{\rm BG} (r) = P_0 \cdot  \rho_{\rm BG} (r) ^{\Gamma};
\label{eq:press_1}
\end{equation}
where 
\begin{equation}
P_0 = \frac{4 \pi G \rho_{\rm s} r_{\rm s}^{2}}{\rho_0^{\Gamma-1}} \frac{\Gamma -1}{\Gamma}, 
\label{eq:normalisation}
\end{equation}
with $G$ the gravitational constant and $\rho_{\rm s}$ the dark matter density at the NFW scale radius $r_{\rm s}$. In this work, we model the polytropic index as 

\begin{equation}
\Gamma = 1 + \frac{(1+x')\ln(1+x')-x'}{(1+3x')\ln(1+x')},
\end{equation}

 \noindent where $x'= c \times \theta_{\rm out}$ and $c$ is the halo concentration. This parametrisation follows the works of \cite{Komatsu&Seljak2001,Bulbul2010,Martizzi2013,Schneider&Teyssier2015,Arico2020}, which assume a gas in hydrostatic equilibrium and polytropic equation of state. By additionally matching the slope of the gas with that of dark matter at large scales, one can avoid having free parameters. Here, we allow more freedom in the profiles to reproduce more closely the results from hydrodynamical simulations across a wide range of halo masses \citep[see e.g.][]{Arico2020b}. 

\subsection{Non-thermal pressure}
Gas within haloes is not solely supported by thermal pressure. Hydrodynamical simulations and (indirect) observations have shown that a considerable amount of non-thermal pressure is expected \citep[see e.g.][]{Shaw2010,Hurier2018,Akino2022}. Taking into account the so-called hydrostatic mass bias, it is possible to reconcile the total mass of the clusters estimated with "direct" measurements (i.e. weak lensing) with those inferred assuming hydrostatic equilibrium from e.g. X-ray or microwave observations. The hydrostatic mass bias is related to the non-thermal pressure support not accounted for, arguably produced by turbulence, cosmic rays, and magnetic fields \citep{Green2020,Ettori&Eckert2022}. 

The non-thermal contribution can be as high as 50\% of the total pressure in the outskirt of the haloes, and therefore impacts significantly tSZ predictions, which are sensitive only to the thermal pressure.

\cite{Shaw2010} have shown with simulated clusters that the total pressure profiles are expected to follow a polytropic equation of state better than the thermal pressure profiles \citep[however, see e.g.][for observational thermal profiles well described by a polytrope]{Ghirardini2019}. In particular, \cite{Green2020} have studied the non-thermal contribution using analytical total pressure profiles \citep{Komatsu&Seljak2001}, non-thermal contribution driven by halo mass assembly \citep{Shi&Kmoatsu2014}, and models for the mass assembly histories \citep{Parkinson2008,vandenBosch2014}. They find that the non-thermal term is roughly universal when expressed as a function of peak height, and they compare their results against hydrodynamical simulations finding a good agreement. 

In this work, we employ Eq.~\ref{eq:press_1} to model the total pressure and then apply the fitting function by \cite{Green2020} to remove the non-thermal contribution. Thus, the thermal pressure profile is 
\begin{equation}
P_{\rm BG, th}(r) = f_{\rm th}(r/r_{\rm 200m}) \, P_{\rm BG} (r),
\end{equation}
where
\begin{equation}
f_{\rm th} (x) = A \left( 1 + e^{-(x/B)^C} \right)  \left( \nu/4.1 \right)^{D / (1+[x/E]^F) }, 
\end{equation}
with the radii $x=r/r_{\rm 200m}$ and peak height $\nu=\nu_{\rm 200m}$ are referred to spheres where the mean density is 200 times larger than the mean matter density of the universe. The best-fitting values found in \cite{Green2020} are $A=0.495$, $B=0.719$, $C=1.417$, $D=-0.166$, $E=0.265$, and $F=-2.116$.

Here, we further assume 
\begin{equation}
A = A_{\rm nth} (1+z)^{\alpha_{\rm nth}}
\end{equation}
with $A_{\rm nth}$ and $\alpha_{\rm nth}$ as free parameters to model possible residual dependences on the amplitude and redshift of the non-thermal contribution, as well as possible departures from the normalisation of the mass-pressure relation assumed in Eq.~\ref{eq:normalisation}.

\subsection{Temperature}

By imposing the ideal gas law, the temperature profile is 

\begin{equation}
T_{\rm BG} (r) = \frac{\mu m_p}{k_{\rm B}} \frac{P_{\rm BG, th}(r)}{\rho_{\rm BG}(r)}, 
\label{eq:ideal_gas}
\end{equation}

 where $m_{\rm p}$ is the proton mass and $k_{\rm B}$ is the Boltzmann constant. We assume for simplicity a fully ionised gas composed of $X=76\%$ hydrogen and $Y=24\%$ helium. Thus, in our case, the mean molecular weight is $\mu = 4/(5X+3) \approx 0.59$. 

In Fig.~\ref{fig:bound_profiles} we show an example of the density, pressure, and temperature profiles in haloes of mass comprised between $10^{13} \, {\it h}^{-1} M_{\odot}$ and $10^{15} \, {\it h}^{-1} M_{\odot}$. In this case, we set the best-fitting parameters to the matter power spectrum of the BAHAMAS simulation \citep{McCarthy2017}. For reference, we over-plot the virial temperature and the ``universal'' pressure profiles from \cite{Arnaud2010}, which have been fitted to massive galaxy clusters of $M \ge 10^{14} \, \Msun$. We observe that, in this specific case, the virial temperature and universal profiles are fairly close to our BCM profiles, especially for massive haloes. We stress that the free parameters of our model impact both the shape and normalisation of the thermodynamical profiles, and can be, in principle, fit to profiles from either simulations or observations.

\subsection{Particle assignment}

Once the profiles have been computed, we assign a given temperature and pressure to each particle belonging to the halo as a function of its distance from the host halo centre, and iterate for all the haloes in a simulation. 
Operationally, we store only the temperature of particles to save memory since the thermal pressure is trivially obtained through the ideal gas law (Eq.~\ref{eq:ideal_gas}). 

For simplicity, we assume that the gas ejected from the haloes by feedback has the same temperature profile as the bound gas, i.e. it thermalises instantaneously. We note that the ejected gas component in massive haloes has typically small mass fractions and low density, thus we expect its contribution to the tSZ to be generally subdominant. 

Next, we assign a temperature to particles that do not belong to any halo. The mass of these gas particles is obtained by weighting the gravity-only particle mass with the cosmic baryon fraction, $m_{\rm g} = \Omega_{\rm b}/\Omega_{\rm m} \times m_{\rm GrO}$. We assume its temperature to be constant and a free parameter of the model, with a fiducial value of $T_{\rm field}=10^6 \, {\rm K}$.
From hydrodynamical simulations and observations, we expect this to be the so-called warm-hot intergalactic medium (WHIM), which has low densities and temperatures between $10^5$ and $10^7 \, {\rm K}$ \citep{Cen&Ostriker1999,Roncarelli2007,deGraaff2019}.

We note that, in previous works, \cite{Mead2020b} fixed the temperature of the gas outside haloes (both the ejected and field components) to $T\approx10^{6.5}\,
{\rm K}$, whereas \cite{Osato&Nagai2022} set $T_{\rm field} =  0 \, {\rm K}$. Assuming $T_{\rm field} =  0 \, {\rm K}$ means to ignore the contribution of the gas around haloes and in filaments, which is subdominant but has been nonetheless detected \citep{VanWaerbeke2014,deGraaff2019}, and can be particularly important in the large-scale cross correlation between total matter and electron pressure. In \cite{Mead2020b}, the ejected gas is added as a linear background in the 2-halo term, because of the technical difficulties of having a gas shell around haloes, since the truncation of dark matter would introduce an anti-correlation with the gas. This procedure, however, likely underestimates the correlation of the ejected gas with their respective haloes, in the transition between 1-halo and 2-halo terms. In our approach this problem is naturally solved.

Finally, we can obtain a total gas mass and temperature fields by summing over ``gas'' particles: $m_{\rm gas} = \sum_i m_i$ and $T_{\rm gas} = \sum_i m_i T_i$. 
We note that, in doing so, the particles will have a smooth and spherical symmetric radial temperature and pressure. On the contrary, the density is not smooth nor spherical symmetric, due to the perturbative nature of the baryonification algorithm (the displacement field is spherical symmetric, but the initial and final particle distribution is not). We expect that the gas clumpiness to weakly impact the tSZ signal, contrary to e.g. the X-ray emissivity \citep{Khedekar2013, Battaglia2015, Ettori2015, Lyskova2023}. We note that the density clumpiness is by construction inherited in the BCM,  and we ignore for now the effects of pressure clumping, leaving its modelling for future works. 

In Fig.~\ref{fig:bar_fields} we show a visualisation of the density field in a baryonified simulation, weighted by gas mass and temperature. We can notice how the temperature-weighted map is more diffuse than the mass, because of the shallower slope of the gas temperature profiles with respect to the density.    

\begin{figure*}
\centering
\includegraphics[width=.49\textwidth]{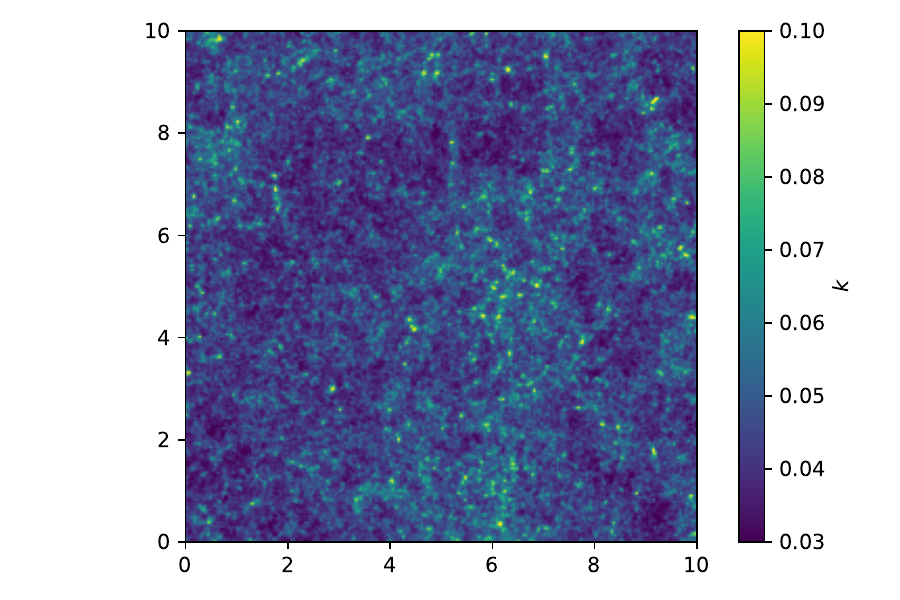}
\includegraphics[width=.49\textwidth]{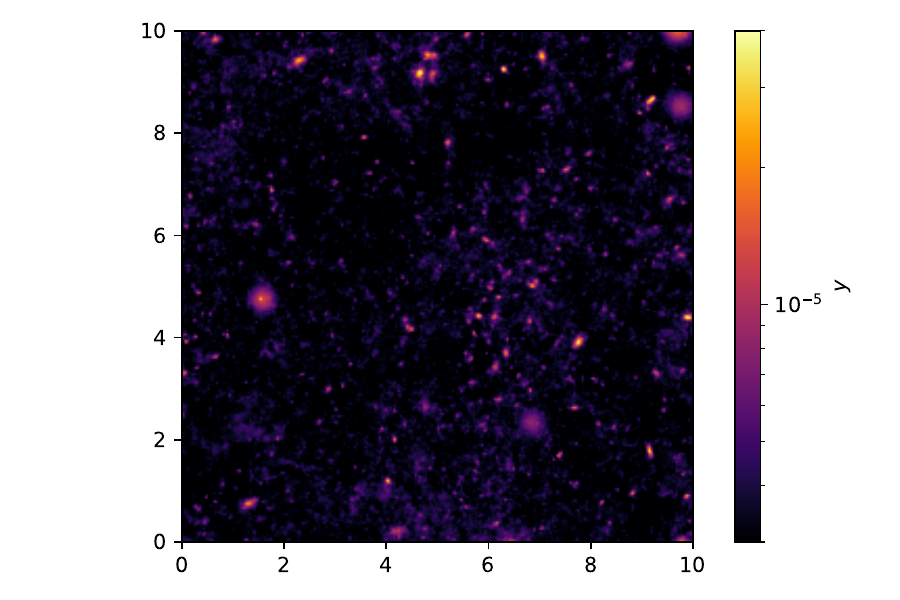}
\caption{Convergence (left panel) and Compton-$y$ maps (right panel) generated by integrating up to $z=1$ a lightcone with a field of view of $10\degree \times 10\degree$ built using a baryonified $N-$ body simulation.}
\label{fig:maps}
\end{figure*}

\section{Thermal Sunyaev-Zel'dovich and convergence}
\label{sec:tsz}
The thermal Sunyaev-Zel'dovich (tSZ) effect is a spectral distortion in the Cosmic Microwave Background (CMB) caused by the inverse Compton scattering of CMB photons with hot electrons. The relative difference in temperature of the photons can be computed as 
\begin{equation}
\frac{\Delta T}{T_{\rm CMB}} = y \left( x \frac{e^x +1}{e^x -1} -4 \right), 
\end{equation}
where $T_{\rm CMB} = 2.726 \, {\rm K}$ and $x = h \nu/(k_{\rm B} T_{\rm CMB} )$ is the dimensionless frequency, $h$ the Planck constant, $k_{\rm B}$ the Boltzmann constant, and $\nu$ the photon frequency.
The parameter $y$ is known as Compton-$y$ parameter,
\begin{equation}
y = \frac{\sigma_{\rm T} k_{\rm B}}{ m_{\rm e} c^2 } \int dl n_{\rm e} (T_{\rm e}-T_{\rm CMB})
\end{equation}
with $m_{\rm e}$, $n_{\rm e}$ and $T_{\rm e}$ the electron rest mass,  density, and temperature, respectively, and $\sigma_{\rm T}$ is the Thomson cross-section. Thus, the Compton-$y$ parameter is the integral of the electron {\it thermal} pressure along the line of sight. 

In the previous section, we have shown how to produce a gas mass and temperature field with baryonification. We can use the mass and temperature to compute a "Compton weight" associated with each gas particle. We follow \cite{Roncarelli2007,McCarthy2014}, and write the weight of the $i$-particle as  
\begin{equation}
\Upsilon_{i} = \frac{\sigma_{\rm T} k_{\rm B} T_i \, m_i }{\mu_{\rm e} m_{\rm p} m_{\rm e} c^2 }
\end{equation}
with $m_{\rm p}$ the rest proton mass, and $c$ the speed of light. The mean molecular weight per electron is assumed to be $\mu_{\rm e} \approx 2/(1+X) \approx 1.14$. 

To predict the angular power spectrum of our Compton-$y$ maps, $C^{yy}_{\ell}$, we can proceed in two different ways: 

i) baryonify the snapshots of a simulation, measure the 3D $y$ power spectrum, $P^{\Upsilon \Upsilon}(k,z)$, and compute $C^{yy}_{\ell}$ by integrating numerically the 3D spectra; \\ 
ii) build a lightcone with the baryonified snapshots, and create a Compton-$y$ map by integrating particles along the line of sight weighted by $\Upsilon$ producing, and measure its angular power spectrum.

The first approach has the flexibility to be applied independently from the specifics of a lightcone, and it could be easily accelerated by emulating the Compton 3D power spectra as a function of redshift. On the other hand, producing lightcones can be useful e.g. to compute higher-order statistics or apply angular masks. Analogously, we can study the WL cosmic convergence, i.e. the apparent distortion of background galaxies caused by the gravitational potential of the LSS, either by integrating the 3D matter power spectrum, or by projecting particles onto a lightcone. 

Either of the two approaches can be combined with a ``cosmology rescaling'', which modifies the cosmology of an $N$-body simulation by perturbing the position and velocity of its particles \citep{A&W2010,Zennaro2019,Contreras2020, Angulo2020}. Explicitly, we can apply the cosmology rescaling and baryonification to a set of snapshots before computing $P^{\Upsilon \Upsilon}(k,z)$ or building the lightcone. This combined approach would allow for a fast exploration of both cosmological and baryonic parameters, while maintaining very accurate predictions.

\subsection{Sky projections}
\label{subsec:integrals}

The angular power spectrum of Compton $y$ can be written as a function of the 3D Compton power spectrum:

\begin{equation}
C^{yy}(\ell) = \int_{0}^{\chi_{\rm H}}  \frac{(1+z(\chi))^2} {\chi^2} P^{\rm \Upsilon\Upsilon} \left( \frac{\ell+1/2}{\chi}, z(\chi) \right) d\chi,
\end{equation}

\noindent where $\chi$ is the comoving distance and $\chi_{\rm H}$ is the comoving distance of the Hubble sphere. 

Analogously, we project the 3D matter power spectrum, $P^{\rm mm}(k,z)$, to get the WL convergence angular power spectrum 

\begin{equation}
C^{k k}(\ell) = \int_{0}^{\chi_{\rm H}}  \frac{g(\chi)^2}{\chi^2} P^{\rm mm} \left( \frac{\ell+1/2}{\chi}, z(\chi) \right) d\chi,
\end{equation}

\noindent where the lensing kernel $g(\chi)$ reads

\begin{equation}
g(\chi) = \frac{3}{2}\Omega_{\rm m} \left( \frac{H_0}{c} \right)^2  \frac{\chi}{a} \int_{z(\chi)}^{z_{\rm H}} n(z') \frac{\chi(z\prime)-\chi(z)}{\chi(z\prime)}dz\prime,
\end{equation}

\noindent with $n(z')$ being the normalised redshift distribution of the sources. We note that this is the convergence power spectrum expected exclusively from the gravitational potential of the LSS, ignoring for simplicity other contributions e.g. the intrinsic alignment of galaxies.  

\subsection{Lightcones: convergence and Compton maps}
\label{subsec:lightcones}

We employ a scheme similar to the onion universe of \cite{Fosalba2008} to create lightcones, where the observer is placed at the centre of the simulation box. We build the lightcone by adding shells of particles at increasingly larger comoving distances from the observer, slicing the simulation snapshots at the correspondent redshift.  

We replicate the box, randomly rotating and mirroring it to avoid the repetition of the same structure along the line of sight. In this way, it is possible to build full-sky maps, as well as maps with a smaller field of view.  

To create the Compton-$y$ maps, we simply integrate along the line of sight the particles weighted by $\Upsilon/A_{\rm p}$, where $A_{\rm p}$ is the physical area subtended by the pixel at the particle distance from the observer. 

Analogously, within the Born and thin lens approximations \citep[see e.g.][]{Petri2017}, we can create convergence maps, $k(\theta)$, by integrating along the line of sight of the excess surface mass density: 
 
\begin{equation}
k(\theta) = \frac{\Sigma (\theta)}{\Sigma_{\rm crit}},
\end{equation}
where the critical surface density is 
\begin{equation}
\Sigma_{\rm crit} = \frac{c^2}{4\pi G} \frac{D_{\rm l}}{D_{\rm s} D_{\rm sl}}, 
\end{equation}

\noindent with $D_{\rm s}$, $D_{\rm l}$, $D_{\rm sl}$ the angular diameter distances of the source, lens, and source-lens planes. In particular, we build 100 equispaced lens planes where we compute $\Sigma (\theta)$ and $\Sigma_{\rm crit}$, assigning particles with a nearest-grid-point scheme to meshes of $2048^2$ pixels. We then sum up the contribution to the convergence for each pixel \citep[see e.g.][]{Fosalba2008,Giocoli2015,Castro2018,Hilbert2020}. 

In Fig.~\ref{fig:maps} we show as an example the convergence and Compton-$y$ maps generated by creating a lightcone with a field of view of $10^{\circ} \times 10^{\circ}$ up to $z=1$. We produce them starting from an $N$-body simulation of $L=512 \hMpc$ of box side and $N=1536^3$ particles, scaling cosmology from {\it TheOne} \citep[see e.g. Table 1 of][]{Pellejero2023} to the nine-year Wilkinson Microwave Anisotropy Probe (WMAP9) \citep{WMAP9} and setting the astrophysics to a BAHAMAS-like feedback \citep{McCarthy2017}. 
The Compton-$y$ map appears clumpier than the convergence, because the tSZ is more sensitive to massive haloes, $y \propto M_{\rm h}^{5/3}$. Furthermore, the tSZ is sensitive to a wide range of redshifts, whereas the convergence has a narrower redshift kernel. For this reason, we employ two different strategies to smooth the maps. For the convergence map, we simply apply a Gaussian smoothing of $\approx0.6'$. For the Compton-$y$ map, we instead adopted a redshift-dependent smoothing. Specifically, we split the lightcone into slices of 50 comoving Mpc along the line of sight, and apply to each slice a Gaussian smoothing proportional to the angular diameter distance of the mean intraparticle distance at the average redshift. Doing so, we have a $\approx 10'$ smoothing for the closest objects, and an increasingly smaller smoothing angles until reaching $\approx 0.1'$ for objects at $\approx 3300 \, {\rm Mpc}$ (z=1). Thus we avoid the strong particle discreteness noise in the Compton-$y$ map coming from clusters close to the observer. 

\begin{figure*}
\centering
\includegraphics[width=.49\textwidth]{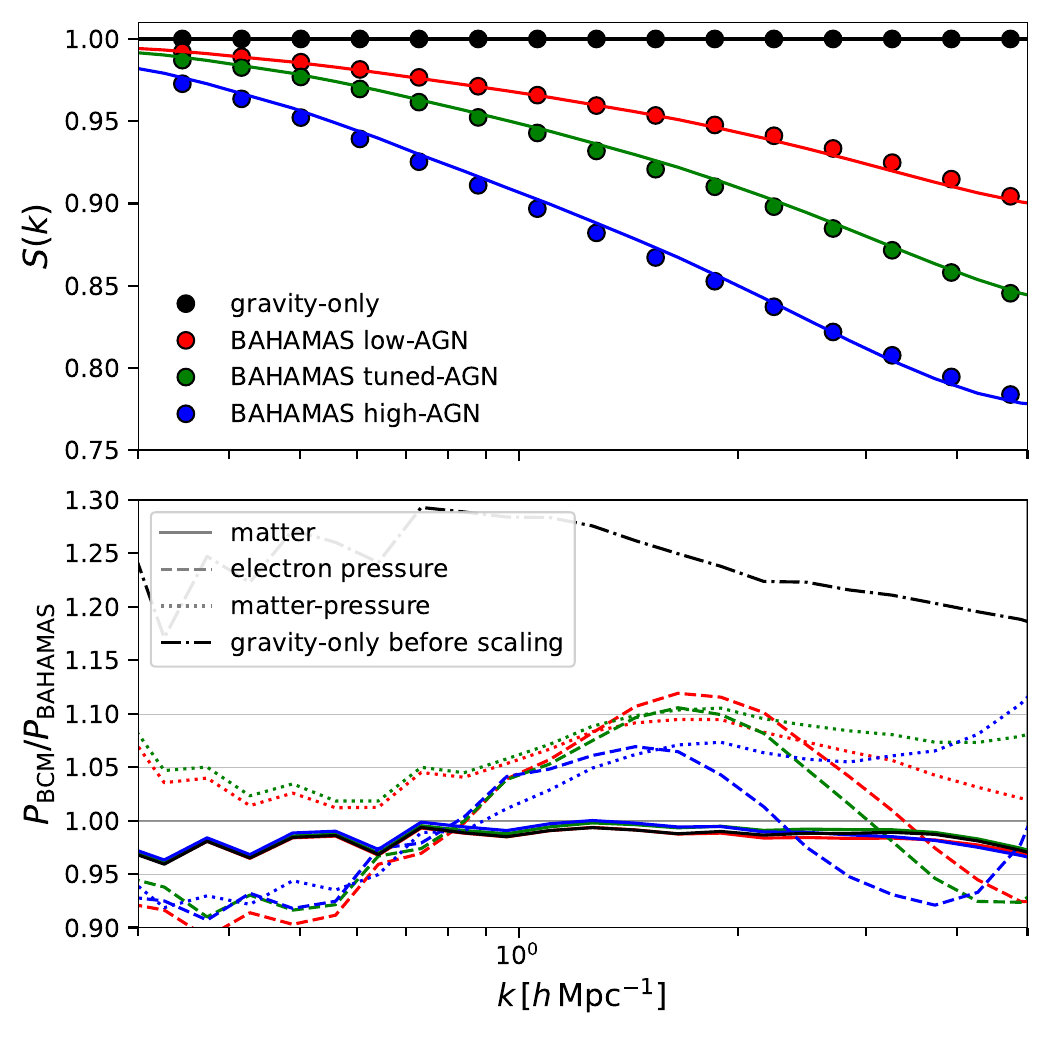}
\includegraphics[width=.49\textwidth]{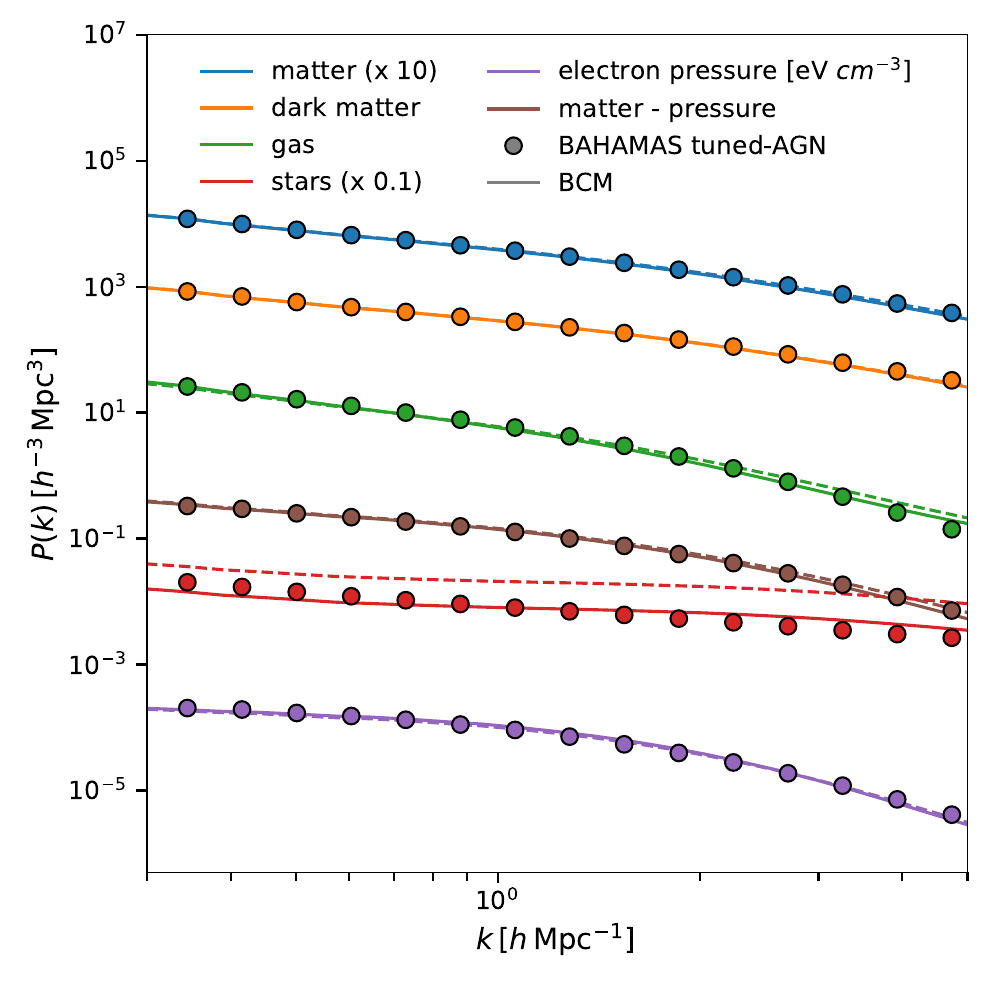}

\caption{{\it Left panel, top}: Matter power spectrum suppression $S(k)$ measured at $z=0$ in three BAHAMAS simulations, with AGN temperature of $T_{\rm AGN}= 10^{7.6}, 10^{7.8}, 10^{8.0}\, {\rm K}$ (red, green, and blue circles, respectively). We show as solid lines the respective predictions of baryonification when simultaneously fitting the $S(k)$, the electron pressure and the matter-electron pressure power spectra of each simulation. {\it Left panel, bottom}: Ratio of power spectra measured in our baryonified simulations to the BAHAMAS ones, for matter (solid lines), electron pressure (dashed lines), and matter-electron pressure cross (dotted lines).  
{\it Right panel}: Power spectra of matter, dark matter, gas, galaxy, electron pressure (in units of $\rm eV \, cm^{-3}$), and matter - electron pressure, measured in BAHAMAS AGN-tuned at $z=0$ (circles, colours as described in the legend). We show the baryonification fits to the total matter and electron pressure auto- and cross-correlation with dashed lines, and a simultaneous fit to all the power spectra with solid lines.}
\label{fig:pks}
\end{figure*}

\section{Comparison with hydrodynamical simulations}
\label{sec:comparison}

In this section, we test if our baryonification approach can reproduce the predictions of hydrodynamical simulations. We compare against BAHAMAS \citep{McCarthy2017,McCarthy2018}, a suite of simulations specifically calibrated with observations of gas fraction in haloes, and, therefore, particularly well suited to study baryonic effects on LSS.

Our comparison will have some intrinsic limitations since i) doing a fit using directly large simulations is computationally demanding, even when using a highly optimised baryonification algorithm; and ii) the BAHAMAS simulation has a box of 400 $\hMpc$, and it is affected by cosmic variance at around 10$\%$ level at $k\approx 0.1 \, \ihMpc$. Focusing on the ratio with respect to the gravity-only matter power spectrum would suppress the cosmic variance of the total matter power spectrum, but this is not trivially true for the electron pressure. In particular, \cite{Mead2020b} found that cosmic variance impacts the BAHAMAS electron pressure power spectrum at the same order of magnitude as the AGN feedback calibration. Therefore, we perform a simple parameter optimisation, leaving to future work the making of emulators of the 3-D Compton power spectrum, to be added to the {\tt BACCOemu} framework \citep{Angulo2020,Arico2020c,Arico2021,Zennaro2023,Pellejero2023}. The emulated Compton power spectrum can be integrated to give the tSZ power spectrum analogously to the matter power spectrum with WL convergence. Moreover, the use of recent large hydrodynamical simulations, e.g. the FLAMINGO suite \citep{flamingo1,flamingo2}, would be a valid option to suppress cosmic variance in future tests.

\begin{figure*}
\centering
\includegraphics[width=.9\textwidth]{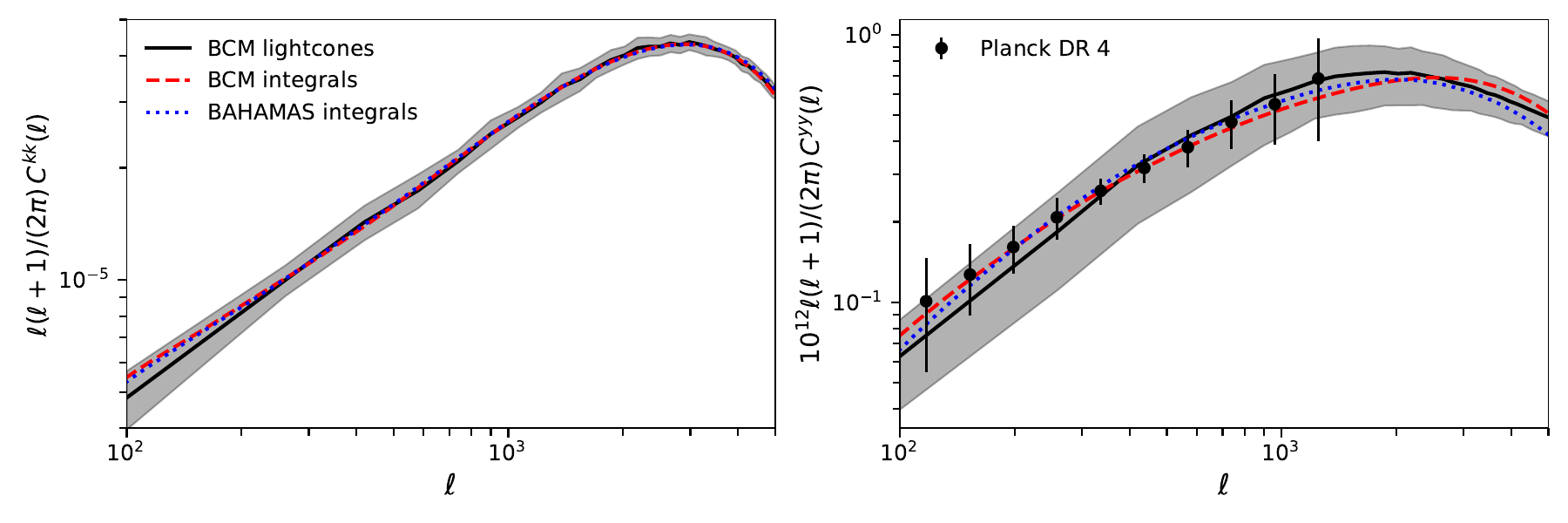}
\caption{Angular power spectra of convergence (left panel) and tSZ (right panel) using the mean of 30 independent baryonified lightcones (black thick line) and integrating numerically the power spectra of the snapshots of our baryonified simulation (blue dotted line). The field of view of each lightcone is 10\degree $\times$ 10\degree, up to $z=1$ (all the sources for convergence are considered to be at $z=1$). The shaded area marks the standard deviation of the 30 lightcone realisations. 
For comparison, we plot the integrals of the 3D power spectra of BAHAMAS $T_{\rm AGN}=10^{7.8}\, {\rm K}$ with blue dotted lines, and the tSZ power spectrum extracted from Plack data release 4 with black circles.}
\label{fig:cls}
\end{figure*}

We note that \cite{Mead2020b} conducted a comparable analysis, using the halo-model-based HMx \citep{Mead2020b, Troster2021} to fit the BAHAMAS total matter auto power spectrum within a few per cent accuracy. However, the fitting accuracy for the total matter-electron pressure cross-power spectrum was limited to $15\%$. These authors also highlighted the challenges in fitting the electron pressure auto power spectrum, as well as the simultaneous fitting of the gas power spectrum and the matter-electron pressure cross power spectrum. They emphasized that difficulties might arise due to either cosmic variance or model limitations, such as the adoption of a simple non-thermal prescription.

In this work, we adopt a more complex non-thermal contribution prescription \citep{Green2020}, embedded in the fully non-linear framework given by the baryonification. We measure the power spectra in two interlaced Fourier grids of $512^3$ cells \citep[folded twice, see e.g.][]{Arico2020b}, and a triangular-shaped clouds mass assignment scheme.

We simultaneously fit the suppression $S(k)$, electron pressure, and mass-electron pressure cross power spectra, considering a 1, 5, and 10\% error in the measurements, respectively, within the scales $k \in [0.3, 5 \, \ihMpc]$ and at $z=0$. We employ a particle swarm algorithm\footnote{\url{https://github.com/hantke/pso_bacco}} for the optimisation, employing 5 particles and 50 steps, varying in logarithmic space a total of 6 free parameters: 4 from the baryonification model by \cite{Arico2020b}, $\log_{10}(M_{\rm c}/(\Msun))$, $\log_{10}\eta$, $\log_{10}\theta_{\rm inn}$, $\log_{10}(M_{\rm 1,z0,cen}/(\Msun))$, plus one to control the temperature of the background gas, $\log_{10}T_{\rm field}$, and one to modulate the amplitude of non-thermal pressure, $\log_{10}A_{\rm nth}$. 
We fix $\log_{10}\beta=-0.45$, $\log_{10}\theta_{\rm out}=0$, $\log_{10}(M_{\rm inn}/(\Msun))=13.5$ to ease the multidimensional optimisation, and, as a further speed up, we initialise the swarm with the best-fitting parameters to the BAHAMAS total matter power spectrum found in \cite{Arico2020c}. After inspecting the BAHAMAS electron pressure at different redshifts, we also set the non-thermal redshift scaling to $\alpha_{\rm nth}$ = 1.5. 

\subsection{3D power spectra}

We show the BAHAMAS suppression in the matter power spectrum, $S(k)$, in the left-top panel of Fig.~\ref{fig:pks}, considering three different AGN feedback scenarios, namely the one with the AGN temperature  $T_{\rm AGN}= 10^{7.6}, 10^{7.8}, 10^{8.0}\, {\rm K}$. Even when fitting them along with the electron pressure auto and cross power spectra $S(k)$, we can reproduce them at the per cent level. In the left-bottom panel of Fig.~\ref{fig:pks}, we report the ratio of the power spectra measured in our simulations to that measured in BAHAMAS. First, we note that the matter power spectrum of our gravity-only simulation run with {\it TheOne} cosmology is $\approx 30\%$ off with respect to the WMAP9-like BAHAMAS. Remarkably, after applying the cosmology scaling (a process without any free parameters involved), the discrepancy is reduced to $\approx 2\%$. Both the electron pressure and matter-electron pressure power spectra are fitted to better than $10\%$ accuracy. We report in Tab.~\ref{tab:model_params} our BCM best-fitting parameters. 

In the right panel of Fig.~\ref{fig:pks}, we show with dash lines our predictions for dark matter, gas, stars, electron pressure, and total-electron pressure power spectra at $z=0$, compared to BAHAMAS $T_{\rm AGN}= 10^{7.8} {\rm K}$. We see that the BCM overestimates the BAHAMAS power spectra of stars at all scales and gas at small scales, which were not explicitly fitted. To check if the model is flexible enough to reproduce them, we refit all the spectra simultaneously, varying the aforementioned free parameters plus the $\alpha_{\rm sat}$ described in  \cite{Arico2020b}, which regulates the mass fraction in satellite galaxies. This parameter has a very weak impact on matter and electron pressure power spectra, but is important to reproduce the stellar power spectrum on large scales. We show in Fig.~\ref{fig:pks} this new fit with solid lines. We find that the gas and stellar power spectra fit are indeed significantly improved. 

\subsection{Angular power spectra}

We employ the best-fitting parameters to the $T_{\rm AGN}= 10^{7.8} \, {\rm K}$ simulation, and build 30 lightcones with a field of view of $10\degree \times 10\degree$ up to $z=1$, following the procedure detailed in \S~\ref{subsec:lightcones}. Each lightcone takes approximately 2.5 hours in 32 cores, including the application of cosmology scaling and baryonification to each of the 24 snapshots used. 

As described in \S~\ref{subsec:lightcones}, we project the lightcones into Compton-$y$ and convergence smoothed maps of $2048\times2048$ pixels, e.g. Fig.~\ref{fig:maps}. We show their power spectra in Fig.~\ref{fig:cls}, and display as a comparison the angular power spectra obtained by integrating numerically the 3-D power spectra of both our baryonified simulation and BAHAMAS, as described in \S~\ref{subsec:integrals}. We measure the power spectra of the convergence and tSZ maps using the routine provided in {\code LENSTOOLS}\footnote{\url{https://github.com/apetri/LensTools/}} \citep{Petri2017}. For comparison, we also display BAHAMAS as dotted lines.

We find that the mean angular power spectrum measured with our 30 lightcone realisations is in excellent agreement with the convergence power spectra integrals, and in good agreement with the tSZ, well within $1\sigma$.   
As expected and discussed in the previous sections, the cosmic variance affects more severely the Compton power spectrum than the convergence. The limited size of our maps, as well as their relatively low number, may cause the differences that we observe. We also note that the binning strategy mildly affects the largest scales shown.    

When comparing our results to BAHAMAS, we find that the convergence is reproduced at $1\%$, whereas the tSZ power spectrum is at $\approx 10\%$. This is compatible with the accuracy we found for the 3D matter and electron pressure power spectra at $z=0$, and thus validates our assumptions of the redshift dependence of baryonic effects. We note that various hydrodynamical simulations might present different redshift dependence, and eventually, this should be tested with observations in a wide range of redshifts. 

As a reference, we overplot the tSZ power spectra obtained from Planck data release 4 data by \cite{Tanimura2022}. The models reproduce well the CMB data, although we remind the reader that we are integrating the tSZ only to $z=1$, and thus we are missing the contribution from higher redshifts. 

Using the BAHAMAS simulations, \cite{McCarthy2014,McCarthy2018} have shown the sensitivity of tSZ to cosmology, and in particular to neutrino masses, as well as to the strength of AGN feedback. The methodology proposed here enables us to explore a much larger parameter space in both cosmology and astrophysics. The BCM is designed to be a rather simple model that can accurately reproduce some statistics of cosmological interest with a few physically-motivated free parameters. However, as we constrain it with an increasing number of multi-wavelength observations and cross-correlations, we should make consistency checks and mock parameter inferences, to test for possible bias introduced by our theoretical assumptions. To this end, we plan to build emulators of the 3D power spectra presented here and perform a more in-depth exploration of the parameter space. Moreover, we warrant further tests of the BCM thermodynamical profiles against both observations and simulations.  

\begin{table}
     \begin{tabular}{l c c c}
    \hline  & \\[-1.5ex]
     $\log_{10} (T_{\rm AGN}/{\rm K})$ &  $7.6$ & $7.8$ & $8.0$ \\ & \\[-1.5ex]
     \hline  & \\[-1.5ex]
     $\log_{10}(M_{\rm c}/(\Msun))$ & 12.90 & 13.69 & 15.2 \\
     $\log_{10}\eta$ & -0.25 & -0.29 & -0.25 \\
     $\log_{10}(M_{\rm 1,z0,cen}/(\Msun))$ & 12.11 & 11.88 & 11.80 \\ 
     $\log_{10}\theta_{\rm inn}$ & -0.28 & -0.26 & -0.20 \\ 
     $\log_{10} (T_{\rm field}/{\rm K})$ & 7.0 & 7.0 & 6.3 \\ 
     $\log_{10}A_{\rm nth}$ & 0.385 & 0.417 & 0.579 \\ & \\[-1.5ex]
     \hline  & \\[-1.5ex]
    \end{tabular}
     \centering
    \caption{Baryonification best-fitting parameters to the matter and electron pressure auto and cross power spectra measured in BAHAMAS at $z=0$, with the AGN feedback strengths specified in the first row.}
\label{tab:model_params}
\end{table}

\section{Conclusions}
\label{sec:conclusions}

In this paper, we have presented an extension of the baryonification method that allows us to model the gas mass, temperature, and pressure in cosmological gravity-only $N$-body simulations. 

After applying the traditional baryonification, we have resampled the matter field with dark matter, gas, and stellar particles. Assuming a polytropic equation of state of the gas inside haloes, we have then assigned a temperature, a pressure, and a Compton weight to each gas particle. 

We have added only 2 extra free parameters, one to regulate the temperature of the gas particles outside haloes and one for the amplitude of the non-thermal pressure contribution. Our novel algorithm enables fast and accurate predictions of the thermal Sunyaev Zel'dovich effect and its cross-correlation with the weak lensing convergence. We have shown that our baryonification, along with the cosmology scaling, can be used to produce lightcones in a wide variety of cosmological and astrophysical scenarios, with a small computational cost (a few CPU hours). 

We have discussed the advantages of our approach with respect to previous works, e.g. \citep{Mead2020b, Osato&Nagai2022}, and in particular its self-consistent modelling of the gas mass and pressure within and outside haloes. 

We have tested our model against the hydrodynamical simulations BAHAMAS \citep{McCarthy2017,McCarthy2018}, finding that it can simultaneously reproduce the total, dark matter, gas, stellar, and electron pressure $3D$ power spectra, as well as the Compton-$y$ and convergence angular power spectra. In particular, we find that we can fit the matter power spectrum suppression at 1\% level and the electron pressure auto and cross-correlation at better than 10\%, down to $k=5 \, \ihMpc$ for different AGN strengths in BAHAMAS. Similarly, we can reproduce the convergence and tSZ power spectra at 1\% and 10\%, respectively, by fixing the redshift dependence of the non-thermal pressure contribution. 

In the future, we plan to build emulators of the 3-D power spectra of the electron pressure and its cross-correlation with the matter density contrast. The same framework can be trivially extended to model fields e.g. kinetic Sunyaev-Zel'dovich, X-ray, and cosmic infrared background. 

We anticipate that this will pave the way for joint analyses of multi-wavelength surveys and cross-correlation of different fields down to non-linear scales, paramount to optimally extracting cosmological and astrophysical information from the planned large-scale structure surveys.  

\section*{Acknowledgements}

We thank Alexander Mead, Tillman Tröster, and Ian McCarthy, for providing the BAHAMAS power spectra data with the related Python functions to read them. We thank Sergio Contreras for the help with the cosmology scaling algorithm, Jens Stücker for the visualisation code used to display the simulations, and Matteo Zennaro, Aurel Schneider, and Daisuke Nagai for useful discussions. We acknowledge the Munich Institute for Astro-, Particle and BioPhysics (MIAPbP) for providing the avenue where some of these discussions have been carried out.
We acknowledge the use of the following software:  {\code BACCO} $\&$ {\code pso BACCO} \citep{Arico2020b}, {\code LENSTOOLS} \citep{Petri2017}, {\code NumPy} \citep{numpy}, {\code SciPy} \citep{scipy}, {\code Matplotlib} \citep{matplotlib}. 

\section*{Data Availability}

The data underlying this article will be shared on reasonable request to the corresponding author.

\bibliographystyle{mnras}
\bibliography{bibliography} 


\end{document}